# Ray-Tracing With a Coherent Ray-Space Hierarchy

Nuno T. Reis[1] and Vasco Costa[1] and João M. Pereira[1]

. [1]Instituto Superior Técnico, University of Lisbon

## Abstract

*We present an algorithm for creating an n-level ray-space hierarchy (RSH) of coherent rays that runs on the GPU. Our algorithm uses rasterization to process the primary rays, then uses those results as the inputs for a RSH, that processes the secondary rays. The RSH algorithm generates bundles of rays; hashes them, according to their attributes; and sorts them. Thus we generate a ray list with adjacent coherent rays. To improve the rendering performance of the RSH vs a more classical approach. In addition the scenes geometry is partitioned into a set of bounding spheres, intersected with the RSH, to further decrease the amount of false ray bundle-primitive intersection tests. We show that our technique notably reduces the amount of ray-primitive intersection tests, required to render an image. In particular, it performs up to 50% better in this metric than other algorithms in this class.*

## 1. Introduction

In Naive Ray-Tracing (RT) each ray is tested against each polygon in the scene, this leads to N x M intersection tests per frame, assuming that we have N rays and M polygons. Performance is thus low, especially with moderately complex scenes due to the sheer amount of intersection tests computed. To optimize this naive approach (and RT in general) there are two common approaches to reduce the number of intersection tests, which are the bottleneck of the algorithm, Object Hierarchies and Spatial Hierarchies. Our work instead focuses on Ray Hierarchies and how to optimize them. This is a less well explored area of the RT domain and one that is complementary to the Object-Spatial Hierarchies.

This paper presents the Coherent Ray-Space Hierarchy (CRSH) algorithm. CRSH builds upon the Ray-Space Hierarchy (RSH) [RAH07] and Ray-Sorting algorithms [GL10]. RSH, described by Roger et al., uses a tree that contains bounding sphere-cones that encompass a local set of rays. The tree is built bottom-up and traversed top-down. Our CRSH algorithm adds Ray-Sorting, described by Garanzha and Loop, to the mix in order to achieve higher efficiency in each tree node and then expands on this basis with whole mesh culling and improved hashing methods.

We hypothesize that improving the coherency of the rays contained within each tree node shall lead to tighter bounding sphere-cones which, in turn, should reduce the amount of ray-geometry intersections. We use specialized ray hashing methods, tuned to the ray types we enumerated (e.g. shadow, reflection and refraction), to further improve the efficiency of the hierarchy. Finally we also introduce whole mesh bounding spheres to reduce even further the number of intersection tests at the top level of the hierarchy. This shallow spherical BVH allows us to further reduce the amount of ray-primitive intersection tests. We note that our technique uses rasterization to determine the primary intersections thus reserving the use of RT for secondaries.

Our main contributions are:

- a compact ray-space hierarchy (RSH) based on ray-indexing and ray-sorting.
- the novel combination of ray-sorting [GL10] with ray-space hierarchy techniques [RAH07] to reduce the amount of ray-primitive intersections.
- culling whole meshes from the RSH prior to performing the final per primitive traversal.

## 2. Background and Related Work

Ray-tracing [Whi80] is a global illumination [RDGK12] technique that is used for the synthesis of realistic images which employs recursive ray-casting [App68].

The ray tracing algorithm casts primary rays from the eye but does not stop there. When the rays intersect geometry they can generate extra secondary rays: e.g. shadow, reflection and refraction rays.

These rays differentiate ray-tracing from the more traditional rasterization algorithm since they naturally allow realistic and complex effects like reflections, refractions and shadows without any need for additional techniques. However this comes with at price. Ray-tracing is computationally expensive. There is extensive research and ongoing to try and optimize it. Much of this research involves creating hierarchies in either the Object or Spatial domains to decrease the number of intersection tests in a divide and conquer fashion.

Object and Spatial Hierarchies help reduce the amount of intersections by reducing the amount of scene geometry required to test by culling many polygons and objects away from the ray paths at once.

Ray-Space Hierarchies, use ray bundles or ray caching mechanisms to achieve the same goal, reduce the number of intersections. Instead of creating hierarchies based on the scenes geometry, they are based on the rays being cast in each frame. This is the approach we decided to take is based on ray bundling but also uses a feature of ray caching, which is ray hashing [AK87] [AK10].

### 2.1. Whitted Ray-Tracing for Dynamic Scenes using a Ray-Space Hierarchy on the GPU

Roger et al. [RAH07] algorithm consists of five steps. The scene is first rasterized, obtaining this way the output information we typically obtain from tracing primary rays using a more traditional ray-tracer. The first batch of secondary rays is generated using this information and then becomes the base for the RSH. The secondary rays are bundled into nodes consisting of a bounding sphere for the rays origin and a bounding cone for the rays direction they create the bottom-level of the hierarchy. Each successive level is created by combining the nodes from earlier levels. Once the top-level is created, the scenes geometry is then intersected with the RSH. Each hit is stored as an integer pair consisting of the triangle identifier and the node identifier. Only these hits will be tested in the lower levels of the hierarchy, effectively reducing the total number of intersection tests with each level. There is an issue however, since the rays aren't sorted in any way, that even at lower levels of the hierarchy the nodes will be too wide and thus intersect too much geometry, which will increase the number of intersection tests necessary. We attempt to solve this problem by sorting the rays prior to creating the RSH.

### 2.2. Fast Ray Sorting and Breadth-First Packet Traversal for GPU Ray Tracing

Garanzha and Loop [GL10] introduced an algorithm using various parallel primitives to adapt the ray-tracing algorithm to the GPU. Their algorithm sorts the generated rays and creates tight-fit frustums on the GPU and then intersects them with a Bounding Volume Hierarchy for the scenes geometry that is built on the CPU. One of the most interesting parts of their work is the fast ray sorting since it is done using parallel primitives. This mitigates the overall time of the operation since it is executed on the GPU. This approach can be merged with Roger et al's [RAH07] algorithm in order to create a more efficient RSH.

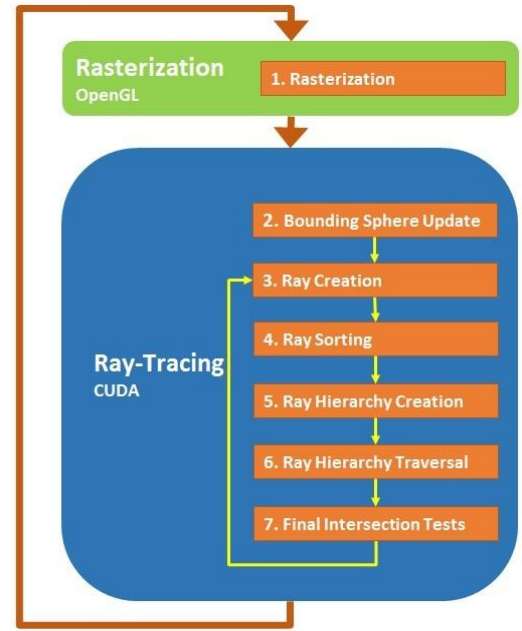

**Figure 1:** *Coherent Ray-Space Hierarchy Overview.*

## 3. Our Algorithm

### 3.1. Overview

Our algorithm is performed in seven steps (see Figure 1). In each frame steps 1 and 2 are executed just once while steps 3 through 7 are executed once per ray batch. Batches can consist of any combination of shadow rays, reflection rays or refraction rays.

### 3.2. Rasterization

Rasterization is performed as a first step. There was in the path an ongoing urban legend that Rasterization and Ray-Tracing are polar opposites and that a choice has to be made between these two techniques. This is not true at all. Although Rasterization solves the rendering problem conversely vs Ray-Tracing (i.e. projecting primitives to the screen, vs projecting rays backwards to the primitives in the scene), one can complement the other. The first set of rays, the primary rays, does not convey any of the global illumination effects that Ray-Tracing is well suited to achieve, such as Shadows, Reflections and Refractions. This means that Rasterization can convey similar visual results to tracing the primary rays, while being extremely faster and optimized in the graphics hardware. Supplementing the Rasterization of primary rays with the Ray-Tracing of secondary rays can get us the benefits from both techniques: the efficiency of Rasterization and the global illumination effects from Ray-Tracing.

In order to combine both techniques the output from the fragment shaders used to rasterize the scene must be more than just the traditional fragment colors. We need to create different render targets according to the information we want to store. In our case we output the fragment diffuse and specular properties, the fragment position and the fragment normal. In our implementation the fragment shader outputs 4 different textures containing four 32 bit floats per pixel each. These textures are generated with OpenGL/GLSL and are then sent to CUDA to create the first level of secondary rays.

### 3.3. Ray-Tracing

#### 3.3.1. Bounding Volume Update

Here we update the object bounding spheres according to the transformations being applied to the object they contain (e.g. translation, scale). Since we only update the center and the radius there is no need to recalculate the bounding spheres in each frame (transformations do not invalidate the bounding spheres).

We pre-compute the minimum bounding sphere of the object meshes, using an implementation based on [Gär99]s algorithm, so there is no impact on render time performance for this computation.

#### 3.3.2. Secondary Ray Generation and Indexing

After the Rasterization step we need to generate the secondary rays. We create an index for each individual ray in order to speed up ray sorting in the following steps. We use a different hashing function for each type of ray (see Figures 2, 3). Since each ray consists of an origin and a direction it would be straightforward to use these two parameters to create our hash. For shadow rays it is sufficient to use the light-index, to which it belongs, and the ray direction. This is doable if we invert the origin of the shadow ray so that it is located at the light source rather than the originating fragment. To further reduce the size of the hash keys we convert the ray direction into spherical coordinates [Pae95] and store both the light index and the spherical coordinates into a 32 bit integer, with the light index having the higher bit-value such that the shadow rays are sorted according to the light source a priori.

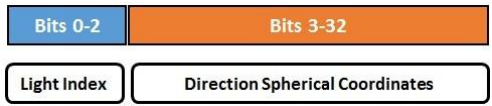

**Figure 2:** *Shadow Ray Hash.*

The Reflection and Refraction rays are also converted to spherical coordinates. However in this case the ray origin is used in the hash, given that these rays are not coherent with regards to the origin, unlike shadow rays.

Once generation is complete, we have an array with the generated secondary rays as well as two arrays with the ray keys (ray hashes) and the ray values (ray position in the ray array) and a final array with head flags which indicate if there is a ray in the corresponding position within the key-value arrays, where we store either a 0 or a 1, indicating if there is a ray or not, respectively. Using the information from the head flags array we then run a trimming operator on the key-value arrays (see Figure 4). This is done by first applying an inclusive scan operator [MG09] on the head flags array, which gives us the number of positions each pair needs to be shifted to the left. This is done in order to trim the arrays [PF05].

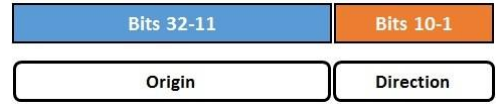

**Figure 3:** *Reflection and Refraction Ray Hash.*

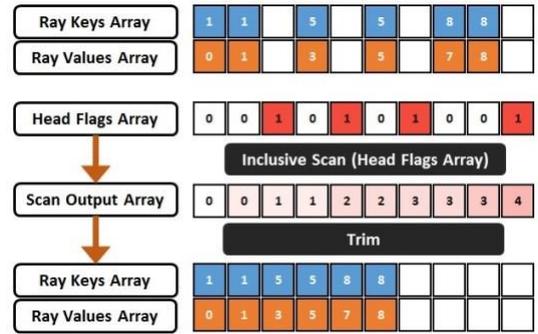

**Figure 4:** *Array Trimming*

#### 3.3.3. Secondary Ray Compression

Here we use a compression-sorting-decompression scheme, expanding on prior work by Garanzha and Loops [GL10]. The compression step exploits the local coherency of rays. Even for secondary rays, the bounces generated by two adjacent rays have a good chance of being coherent. This can result in the same hash value for both bounces. Given this information, we compress the ray key-value pairs into chunks, minimizing the number of pairs that need to be sorted. To compress the pairs we utilize a head flags array with the same size as the key-value pair array, initializing it with 0s in every position and inserting 1s into positions in which the key (hash) of the corresponding pair differs from the previous pair. After populating the head flags array we apply an inclusive scan operator on it [MG09]. By combining the head flags array with the scan output array we create the chunk keys, base and size arrays, which contain the hash, starting index and size of the corresponding chunks (see Figure 5). The chunk keys are represented in different colors at the image below. The chunk base array represents the original position of the first ray in the chunk while the chunk size array represents the size of the chunk, needed for the ray array decompression.

#### 3.3.4. Secondary Ray Sorting

After ray compression we have an array of chunks with the information required to reconstruct the initial rays array. So we can begin the actual sorting. We radix sort [MG10] the chunks array according to the chunk keys.

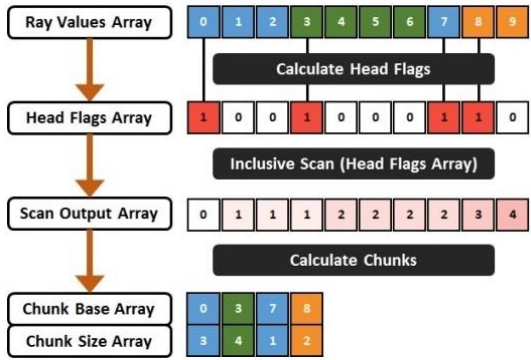

**Figure 5:** *Ray Compression into Chunks*

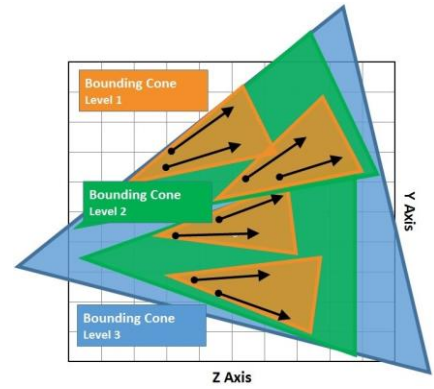

**Figure 7:** *Bounding Cone - 2D View*

### 3.3.5. Secondary Ray Decompression

Decompression works by creating a skeleton array. This skeleton array is similar to the head flag arrays we created before except that it contains the size of the sorted chunks. Next we apply an exclusive scan operator on the skeleton array. This will give us the positions of the chunks starting positions on the sorted key and value arrays. After creating these two arrays for each position in the scan array we fill the sorted ray array. We start in the position indicated in the scan array and finish after filling the number of rays contained within the corresponding chunk.

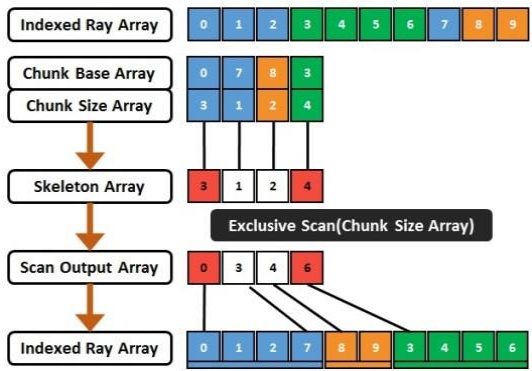

**Figure 6:** *Ray Decompression from Chunks*

### 3.3.6. Hierarchy Creation

With the sorted rays we can now create the actual hierarchy. Since the rays are now sorted coherently the hierarchy will be much tighter in its lower levels, giving us a smaller number of intersection candidates as we traverse further down the hierarchy. Each node in the hierarchy is represented by a sphere and a cone (see Figures 7, 8).

The sphere contains all the nodes ray origins while the cone contain the rays themselves (see Figure 9). This structure is stored using eight floats: the sphere center and radius (four floats) and the cone direction and spread angle (four floats). The construction of the hierarchy is done in a bottom-up fashion. Thus we start with the leaves, with spheres of radius 0 and a cone with spread angle equal to 0. These leaves correspond to the sorted rays. The upper levels of the hierarchy are created by calculating the union of the child nodes. The number of children combined in each node can also be parametrized.

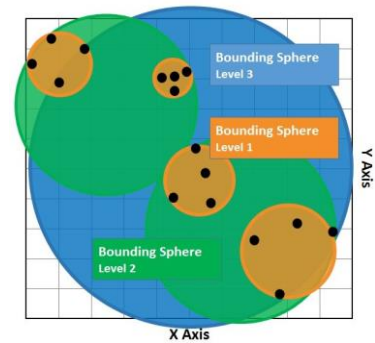

**Figure 8:** *Bounding Sphere - 2D View*

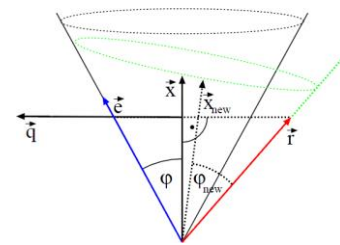

**Figure 9:** *Cone-Ray Union - 2D View. courtesy of [Szé06].*

For the first level of nodes we use the formulas below to create compact cones [Szé06].

$$\rightarrow q = \frac{(\rightarrow x_i \hookleftarrow x_j) \hookleftarrow x_k \hookleftarrow x_l}{|(\rightarrow x_i \hookleftarrow x_j) \hookleftarrow x_k \hookleftarrow x_l|} \qquad (1)$$

$$\rightarrow e = \rightarrow x_i \cdot \cos(\phi) + \rightarrow q \cdot \sin\phi \qquad (2)$$

$$\rightarrow x_{new} = \frac{\rightarrow e + \rightarrow r}{|\rightarrow e + \rightarrow r|} \qquad (3)$$

$$\cos\phi_{new} = \rightarrow x_{new} \cdot \rightarrow r \qquad (4)$$

For the remaining levels we use the following formulas to combining cones:

$$\rightarrow x_{new} = \frac{\rightarrow x_1 + \rightarrow x_2}{|\rightarrow x_1 + \rightarrow x_2|} \qquad (5)$$

$$\cos\phi_{new} = \frac{\arccos(\rightarrow x_1 + \rightarrow x_2)}{2} + \max(\phi_1, \phi_2) \qquad (6)$$

Finally for the union of spheres we use this formula:

$$center_{new} = \frac{center_1 + center_2}{2} \qquad (7)$$

$$radius_{new} = \frac{|center_2 - center_1|}{2} + \max(radius_1, radius_2) \qquad (8)$$

Each ray needs to know the pixel that it corresponds to. After the sorting step rays are not in their original order. We need a way to map rays back to the screen pixels. Since the hierarchy is not tied directly to the geometry positions in the scene it does not matter for hierarchy creation whether the scene is dynamic or static. The only thing which does matter is the number of bounces of each ray, meaning that if there are more pixels occupied in the screen, the hierarchy will have more nodes.

Roger et al. [RAH07] also noted that some nodes might become too large as we travel higher up into the hierarchy. To mitigate this problem we decided to limit the number of levels generated and subsequently the number of levels traversed. Since rays are sorted before this step, there is much higher coherency between rays in the lower levels. If we focus on these rays and ignore the higher levels of the hierarchy we will have better results (this will be demonstrated later in the evaluation section). There is a possibility that we might end up having more local intersection tests but since the nodes in the higher levels of the hierarchy are quite large, we would most likely end up having intersections with every single triangle. Thus having no real gain from calculating intersections on these higher level nodes to begin with.

### 3.3.7. Hierarchy Traversal

Once we have an hierarchy we can traverse it. Prior to traversing the hierarchy we compute the bounding spheres for each object in the scene using Bernd Gartners algorithm [Gär99].

For the top level of the hierarchy we intersect the hierarchy nodes with the bounding spheres to cull intersections further. Finally we traverse the hierarchy in a top-down order, intersecting each node with the scenes geometry. Since the top level nodes of the hierarchy fully contain the bottom level nodes, triangles rejected at the top levels will not be tested again in the bottom level nodes. Let us say we start traversing the hierarchy with the root node. If a certain triangle does not intersect the root node then this means that that specific triangle will not intersect any of its children. Since it is the root node, it also means that no ray in the scene will intersect it so we do not have to further test it for intersections. After traversing each level of the hierarchy we store the intersection information in an array so that the child nodes will know the sets of triangles they have to compute intersections against. The intersection tests being run at this stage are only coarse grained tests. They use the triangles bounding spheres since we will have to do the actual intersection tests in the final stage anyway. The intersection tests are being run in a parallel manner so there is an issue regarding empty spaces in the textures that contain the intersection information. These arrays need to be trimmed using the same procedure that we used after the ray generation. These hits are stored as an int32 in which the first 18 bits store the node id and the last 14 bits store the triangle id. This is not a problem for larger scenes since those processed in triangle batches. Each hit only needs to store the maximum number of triangles per batch.

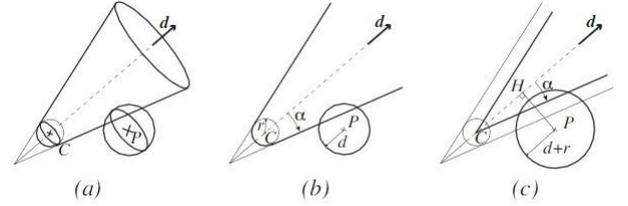

**Figure 10:** *Cone-Ray Union - 2D View. courtesy of [RAH07].*

To calculate the intersection between the node, which is composed of the union of a sphere and a cone, we simplify the problem by enlarging the triangles bounding sphere [Eri04] and reducing the cones size (see Figure 10). The original formula for cone-sphere intersections was described in the Amanatides paper [Ama84]. The current formula, which expands on the work of Amanatides, was described by Roger et al. [RAH07].

$$result = |C - H| \times \tan\alpha + \frac{d + r}{\cos\alpha} \geq |P - H| \qquad (9)$$

### 3.3.8. Final Intersection Tests

After traversing the hierarchy we have an array of node id and triangle id pairs. The candidates for the local intersection tests [Möl97]. In this final step all that remains is to find out which is the closest intersected triangle for each ray and accumulate shading. Depending on the depth that we want for the algorithm we might need to output another set of secondary rays. Since the algorithm is generic, all that is necessary for this is to output these rays onto the ray array that we used initially and continue from the ray-sorting step.

## 4. Evaluation

### 4.1. Test Methodology

We implemented our CRSH algorithm in OpenGL/C++ and CUDA/C++ then compared it with our implementation of RAH [RAH07] over the same architecture. We map our algorithm onto the GPU, parallelizing it there fully. We achieve this mainly by the use of parallel primitives, like prefix sums [Ble90]. We used the CUB [MG09] [MG10] library to perform parallel radix sorts and prefix sums.

We measure the amount of intersections, including misses and hits, to evaluate ray hierarchy algorithms proficiency at reducing the amount of ray-primitive intersection tests required to render an image. All scenes were rendered at $512 \times 512$ resolution using a hierarchy depth of 2 and a node subdivision of 8 (each upper level node in the hierarchy consists of the combination of 8 nodes from the level directly below).

The test information was collected using a NVIDIA GeForce GTX TITAN GPU with 6 GB of RAM. Our algorithm is completely executed on the GPU (including hierarchy construction and traversal) so the CPU has no impact on the test results.

### 4.2. Test Scenes

We used three different scenes, OFFICE, CORNELL and SPONZA.

The OFFICE scene (36K triangles) is representative of interior design applications. It is divided into several submeshes therefore it adapts very well to our bounding volume scheme. For this scene the emphasis was on testing shadow rays.

We selected CORNELL (790 triangles). as it is representative of highly reflective scenes. It consists an object surrounded by six mirrors. On this scene we focused on testing reflection rays although it also features shadow rays in it.

SPONZA (66K triangles), much like OFFICE, is representative of architectural scenes. For this scene the emphasis was also on testing shadow rays but for scenes that do not conform with our bounding volume scheme. This scene does not adapt well to our scheme as is not divided into submeshes.

### 4.3. Test Results and Discussion

#### 4.3.1. Intersection Results

We hypothesized that our more coherent RSH hierarchy needs to compute fewer intersection results to render a scene. We expect more expressive results for shadow rays. As shadow rays have low divergence the sorting step should create a more coherent RSH than for reflection rays. In addition, our hierarchy should also be more coherent with reflection rays than one based on the RAH algorithm due to the hashing we use. However, by the very nature of reflection rays they will never be as coherent as shadow rays resulting in a lower quality hierarchy.

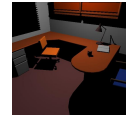

| OFFICE | LEVEL 2 | LEVEL 1 |
|---|---|---|
| *RAH Algorithm* | | |
| # SH INTERSECTIONS | 142726748 | 202025920 |
| # SH MISSES | 117473508 | 186409563 |
| # SH HITS | 25253240 | 15616357 |
| *Our Algorithm* | | |
| # SH INTERSECTIONS | 11559388 | 85572416 |
| # SH MISSES | 862836 | 76455023 |
| # SH HITS | 10696552 | 9117393 |

**Table 1:** OFFICE *rendering performance (251546 shadow rays).*

Our initial expectations for Office were to get a much lower number of intersection tests with our algorithm than with RAH. The scene is a good fit to our bounding volume scheme and our highly coherent shadow ray hierarchy. Results confirm (see Table 1) our initial expectations: we compute 63,79% less intersections than RAH on this scene. 98,14% less than a brute force approach.

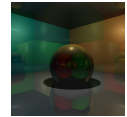

| OFFICE | LEVEL 2 | LEVEL 1 |
|---|---|---|
| *RAH Algorithm* | | |
| # SH INTERSECTIONS | 2995344 | 5167632 |
| # SH MISSES | 2349390 | 3606077 |
| # SH HITS | 645954 | 1561555 |
| | | |
| # RE INTERSECTIONS | 6488064 | 17802296 |
| # RE MISSES | 4262777 | 14311812 |
| # RE HITS | 2225287 | 3490484 |
| *Our Algorithm* | | |
| # SH INTERSECTIONS | 750384 | 2375896 |
| # SH MISSES | 453397 | 981144 |
| # SH HITS | 296987 | 1394752 |
| | | |
| # RE INTERSECTIONS | 2737872 | 10269256 |
| # RE MISSES | 1454215 | 6983410 |
| # RE HITS | 1283657 | 3285846 |

**Table 2:** CORNELL *rendering performance (184717 shadow & 524288 reflection rays).*

For the Cornell scene we focused primarily on the reflection rays which are much more incoherent than shadow rays so we expected results to be less positive than with the Office scene. We compute 26,47% less intersections overall (shadow and reflection rays combined) than the RAH algorithm and 91,17% less than the brute force approach (see Table 2).

The final scene, Sponza, is a whole mesh. We did not employ object subdivision in this scene. Hence, we expected worse results than with Office since we would only get the benefit of the shadow ray hierarchy and none from the bounding volume scheme.

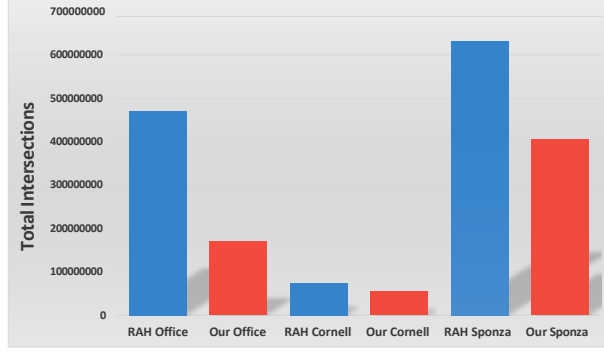

| ALGORITHM | OFFICE | | CORNELL | | SPONZA | |
|---|---|---|---|---|---|---|
| | TOTAL # ISECT | RELATIVE % | TOTAL # ISECT | RELATIVE % | TOTAL # ISECT | RELATIVE % |
| *Brute Force* | 9133132168 | 100% | 606911976 | 100% | 17058578850 | 100% |
| *RAH Algorithm* | 469683524 | 5.14% | 72869648 | 12.01% | 632408864 | 3.71% |
| *Our Algorithm* | **170070948** | **1.86%** | **53578192** | **8.83%** | **405619896** | **2.38%** |

**Table 4:** OFFICE *(251546 shadow rays)*, CORNELL *(184717 shadow & 524288 reflection rays)*, SPONZA *(256713 shadow rays) rendering performance.*

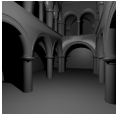

| OFFICE | LEVEL 2 | LEVEL 1 |
|---|---|---|
| *RAH Algorithm* | | |
| # SH INTERSECTIONS | 266597400 | 261494752 |
| # SH MISSES | 233910556 | 248455163 |
| # SH HITS | 32686844 | 13039589 |
| | | |
| *Our Algorithm* | | |
| # SH INTERSECTIONS | 266597400 | 62665496 |
| # SH MISSES | 258764213 | 53120871 |
| # SH HITS | 7833187 | 9544625 |

**Table 3:** SPONZA *rendering performance (256713 shadow rays).*

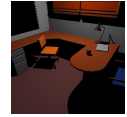

| OFFICE | TOTAL TIME (MS) | RELATIVE TIME (%) |
|---|---|---|
| *RAH Algorithm (618 ms per Frame)* | | |
| # RAY CREATION | 40,240 | 0,11% |
| # RAY COMPRESSION | 0,000 | 0,00% |
| # RAY SORTING | 0,000 | 0,00% |
| # RAY DECOMPRESSION | 0,000 | 0,00% |
| # HIERARCHY CREATION | 122,165 | 0,34% |
| # HIERARCHY TRAVERSAL | 30394,169 | 84,77% |
| # FINAL INTERSECTION TESTS | 4428,185 | 12,35% |
| | | |
| *Our Algorithm (316 ms per Frame)* | | |
| # RAY CREATION | 40,433 | 0,22% |
| # RAY COMPRESSION | 16,651 | 0,09% |
| # RAY SORTING | 11,822 | 0,06% |
| # RAY DECOMPRESSION | 99,721 | 0,54% |
| # HIERARCHY CREATION | 130,626 | 0,71% |
| # HIERARCHY TRAVERSAL | 13918,797 | 75,76% |
| # FINAL INTERSECTION TESTS | 3355,308 | 18,26% |

**Table 5:** OFFICE *rendering time.*

We compute 35,86% less intersection tests than RAH and 97,62% than the brute force approach. Since there is no object mesh culling for this scene the results are not as good as they were with the Office scene but we still manage to outperform RAH even without using an integral part of our algorithm (see Table 3).

### 4.3.2. Intersection Results

These tests were run over the course of 58 frames and the results for each phase are the average of these 58 frames.

For this scene we can see that the major time consuming steps are in fact the hierarchy traversal and the final intersection tests. The biggest difference between our algorithm and the RAH algorithm resides in the time spent traversing the hierarchy. While our algorithm only needs 13918 milliseconds, the RAH algorithm requires 30394 milliseconds, which is about 215% more. This increased time traversing the hierarchy amounts to about 100% more time to render each frame (see Table 5).

Much like Office, the Cornell scene takes most of its rendering time traversing the hierarchy and calculating the final intersection tests. However due to the lower geometric complexity of the scene the absolute values aren't as high. The traversal takes 2717 milliseconds for the RAH algorithm and 1700 for our algorithm, which is a significant reduction (see Table 6).

Finally for the Sponza scene we still see a similar relative time being spent in the traversal of the hierarchy as in the previous scenes. Even though the Sponza scene isn't subdivided into separate object meshes, we still manage to outperform RAH while traversing the hierarchy (see Table 7).

## 5. Conclusions and Future Work

Our paper introduced a new algorithm that creates an efficient Ray-Space Hierarchy which greatly reduces the number of intersection tests necessary to ray-trace a scene due to its improved coherency and shallow bounding volume hierarchy.

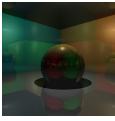

| CORNELL | TOTAL TIME (MS) | RELATIVE TIME (%) |
|---|---|---|
| *RAH Algorithm (93 ms per Frame)* | | |
| # RAY CREATION | 125,731 | 2,16% |
| # RAY COMPRESSION | 0,000 | 0,00% |
| # RAY SORTING | 0,000 | 0,00% |
| # RAY DECOMPRESSION | 0,000 | 0,00% |
| # HIERARCHY CREATION | 366,984 | 6,29% |
| # HIERARCHY TRAVERSAL | 2717,309 | 46,59% |
| # FINAL INTERSECTION TESTS | 2208,936 | 37,88% |
| | | |
| *Our Algorithm (100 ms per Frame)* | | |
| # RAY CREATION | 128,134 | 2,38% |
| # RAY COMPRESSION | 53,916 | 1,00% |
| # RAY SORTING | 50,970 | 0,95% |
| # RAY DECOMPRESSION | 251,946 | 4,68% |
| # HIERARCHY CREATION | 407,090 | 7,57% |
| # HIERARCHY TRAVERSAL | 1776,750 | 33,03% |
| # FINAL INTERSECTION TESTS | 2175,057 | 40,43% |

**Table 6:** CORNELL *rendering time.*

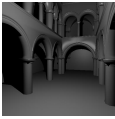

| OFFICE | TOTAL TIME (MS) | RELATIVE TIME (%) |
|---|---|---|
| *RAH Algorithm* | | |
| # RAY CREATION | 41,215 | 0,07% |
| # RAY COMPRESSION | 0,000 | 0,00% |
| # RAY SORTING | 0,000 | 0,00% |
| # RAY DECOMPRESSION | 0,000 | 0,00% |
| # HIERARCHY CREATION | 126,499 | 0,23% |
| # HIERARCHY TRAVERSAL | 49537,627 | 89,25% |
| # FINAL INTERSECTION TESTS | 4995,089 | 9,00% |
| | | |
| *Our Algorithm* | | |
| # RAY CREATION | 41,500 | 0,08% |
| # RAY COMPRESSION | 17,151 | 0,03% |
| # RAY SORTING | 12,696 | 0,02% |
| # RAY DECOMPRESSION | 98,617 | 0,19% |
| # HIERARCHY CREATION | 137,835 | 0,26% |
| # HIERARCHY TRAVERSAL | 46867,912 | 89,65% |
| # FINAL INTERSECTION TESTS | 4367,193 | 8,35% |

**Table 7:** SPONZA *rendering time.*

We achieved our goal of reducing the number of intersection tests using a Ray-Space Hierarchy. This technique is orthogonal to the use of both Object and Space Hierarchies. They can all be be used together in the same application to obtain even better results. Our results show that we can expect a reduction in computed intersections of 50% for shadow rays and 25% for reflection rays compared to previous state of the art RSHs.

However, there is still room for improvement, namely in two areas: The first area of improvement concerns the ray hashing functions. Since the hash determines how rays are sorted, the hierarchy will improve if we manage to enhance the ray classification accuracy (i.e. ray spatial coherency).

The second area of improvement relates to the object bounding-volumes. We used spherical bounding volumes in this paper and a shallow object hierarchy. In the future we aim to also combine our coherent ray hierarchy with a deeper object hierarchy that will further reduce the number of ray-primitive intersections (e.g. [BO04]).

**Appendix**

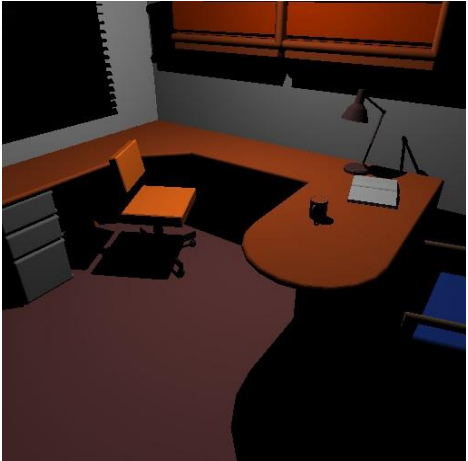

OFFICE

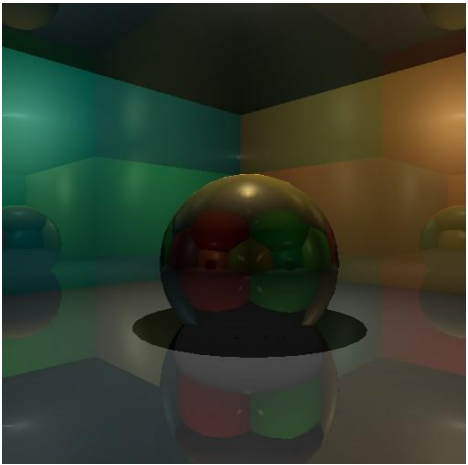

CORNELL

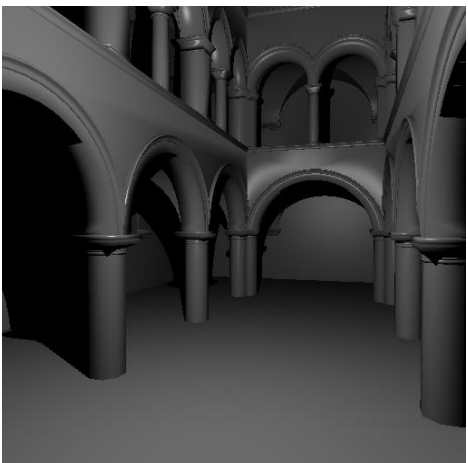

SPONZA